\documentclass[useAMS,usenatbib]{mn2e}
\pdfoutput=0

\usepackage{graphicx}
\usepackage{aasmacros}
\usepackage{color}
\usepackage{epsfig}

\def\msol{M$_\odot$}

\def\om{\Omega_m}

\def\omb{\Omega_b}
\def\s8{\sigma_8}

\def\lcdm{$\Lambda$CDM}

\def\x2{$\chi^2$}

\def\NNm1{\langle N(N-1) \rangle}

\def\slogm{\sigma_{{\rm log}M}}
\def\m_star{M_\ast}

\def\lcdm{$\Lambda$CDM}
\def\slogm{\sigma_{{\rm log}M}}
\def\om{\Omega_m}

\def\omb{\Omega_b}
\def\s8{\sigma_8}

\def\x2{$\chi^2$}

\def\NNm1{\langle N(N-1) \rangle}

\def\p0{P_0(r)}

\def\fq{f_{\rm Q}}

\def\mgal{M_\ast}

\def\mhalo{M_{h}}

\def\slogm{\sigma_{\log M\ast}}

\def\fcon{f_{\rm con}}
\def\mhcrit{M_h^{\rm crit}}
\def\mgcrit{M_\ast^{\rm crit}}
\def\zcrit{z_{\rm crit}}
\def\ratio{\mgal/\mhalo}
\def\ratiocrit{\mgal/\mhalo^{\rm crit}}

\def\zz{Z/Z_\odot}
\def\sigstoch{\sigma_{\rm stoch}}

\title[Testing Galaxy Quenching with Scatter]{
Testing Galaxy Quenching Theories with Scatter \\in the Stellar to
Halo Mass Relation}

\author[Tinker]{Jeremy L. Tinker$^1$\\
  $^1$Center for Cosmology and Particle Physics, Department of Physics, New York University, New York, NY}
\begin{document}


\pagerange{\pageref{firstpage}--\pageref{lastpage}} \pubyear{2016}

\maketitle

\label{firstpage}

\begin{abstract}

  We use the scatter in the stellar-to-halo mass relation to constrain
  galaxy evolution models. If the efficiency of converting accreted baryons
  into stars varies with time, halos of the same present-day mass but
  different formation histories will have different $z=0$ galaxy
  stellar mass. This is one of the sources of scatter in stellar mass
  at fixed halo mass, $\slogm$. For massive halos that undergo rapid
  quenching of star formation at $z\sim 2$, different mechanisms that
  trigger this quenching yield different values of $\slogm$. We use
  this framework to test various models in which quenching begins
  after a galaxy crosses a threshold in one of the following physical
  quantities: redshift, halo mass, stellar mass, and stellar-to-halo
  mass ratio. Our model is highly idealized, with other sources of
  scatter likely to arise as more physics is included. Thus, our test
  is whether a model can produce scatter lower than observational
  bounds, leaving room for other sources. Recent measurements find
  $\slogm=0.16$ dex for $10^{11}$ \msol\ galaxies. Under the
  assumption that the threshold is constant with time, such a low
  value of $\slogm$ rules out all of these models with the exception
  of quenching by a stellar mass treshold. 
  Most physical quantities, such as metallicity, will increase scatter
  if they are uncorrelated with halo formation history. Thus, to
  decrease the scatter of a given model, galaxy properties would
  correlate tightly with formation history, creating testable
  predictions for their clustering. Understanding why $\slogm$ is so
  small may be key to understanding the physics of galaxy formation.

\end{abstract}


\section{Introduction}

In its simplest form, abundance matching connects galaxies with dark
matter halos by the rank-order of both objects: the $N$th most massive
galaxy resides in the $N$th most massive halo.  The success of this
paradigm rests on the assumption all halos of mass $\mhalo$ have
galaxies with mass stellar $\mgal$ inside them, regardless of the
formation history of each halo. Any scatter in the relation is put in
by-hand, post-facto. In essence, abundance matching rests on the idea
that galaxy formation is a `path-independent' process. Using the mean
growth of halos, combined with measurements of the galaxy stellar mass
function at various redshifts, one can use abundance matching to
determine the average path of stellar mass growth in bins of halo mass
(\citealt{conroy_wechsler:09, behroozi_etal:13_letter,
  behroozi_etal:13}, hereafter B13, \citealt{moster_etal:13}). From
this, one can show the efficiency of converting accreted baryons into
stars, $\fcon$. For massive halos---those whith $z=0$ masses of
$10^{13}$ \msol, which will be the focus of this paper---this function
monotonically increases with cosmic time up until it peaks at $z\sim
2$, whereupon it turns over and galaxy growth quickly stalls. These
results are in agreement with analyses of the stellar populations
of massive galaxies, which imply rapid galaxy growth at high redshift
with limited growth after $z\sim 2$ (e.g., \citealt{thomas_etal:05}).

These results reflect the {\it average} formation history of galaxies
within halos.  However, dark matter halos of fixed mass can have
widely varying formation histories. Two halos with a present-day mass
of $10^{13}$ \msol\ can differ by a factor of five at 1-$\sigma$ at
$z=3$ (e.g., \citealt{wechsler_etal:02}).  Any dependence of the
baryonic conversion efficiency, $\fcon$, with redshift---either
explicitly, or implicitly through a dependence on $\mhalo(z)$,
$\mgal(z)$, or other quantity---with break the path-independence of
galaxy formation. Two halos with the same $z=0$ dark matter mass will
{\it not} have the same mass galaxy in them. The distribution of halo
formation histories is then one of the prime sources of scatter in the
stellar mass to halo mass relation.

In this paper we will use the measurements of the scatter in stellar
mass at fixed halo mass, $\slogm$, to put constraints on how $\fcon$
can vary with time for massive galaxies of present-day stellar mass
$\mgal\approx 10^{11}$ \msol. These galaxies form in halos of
$\mhalo\approx 10^{13}$ \msol (B13, \citealt{moster_etal:13}). We
focus on massive galaxies for two reasons: first, these galaxies are
nearly uniformly quiescent (\citealt{chen_etal:12, reid_etal:16}), thus
the process that quenches star formation has already occurred in these
halos. As we will show, because this process of quenching must occur
over a short time span, it causes an extreme break of the
path-independence of galaxy formation, and thus has a strong impact on
$\slogm$. Second, although these galaxies are quite massive, abundance
matching informs us that the buildup of stellar mass within these
halos is due to in-situ star formation, and not by merging. B13 and
\cite{moster_etal:13} both find that the fraction of stellar mass from
in-situ growth in these halos if 90\% at $z=0$ and 95\% at $z=0.5$,
which is the redshift of the Baryon Oscillation Spectroscopic Survey
(BOSS; \citealt{dawson_etal:13}) galaxy sample from which we will take
observations of $\slogm$. Thus, the dominant source of scatter in
stellar masses within these halos is star formation and not merging,
which can dominate $\slogm$ in higher mass halos
(\citealt{gu_etal:16}).

Using the clustering and abundance of BOSS galaxies,
\cite{tinker_etal:16_boss} found $\slogm$ for $10^{13}$ \msol\ halos
to be 0.16 dex. This value removes statistical errors from the stellar
mass estimates, but does not remove systematic random errors incurred
from the stellar mass estimation method itself. This value is in good
agreement with other measurements of $\slogm$ from other galaxy
samples at lower redshifts (\citealt{more_etal:11, reddick_etal:13,
  zu_mandelbaum:16}).  Although this is an upper limit, 0.16 dex is a
shockingly low value for a quantity---stellar mass---that is
influenced by a series of disparate processes, all with their own
intrinsic distributions, such as metallicity, AGN feedback, supernovae
feedback and winds, gas-rich merging, and the different baryonic
accretion rates that different halos experience. The models we present
in this paper are highly simplified, incorporating none of the
physical effects just listed. Thus, our test for whether a model for
quenching is valid is whether it can yield a value of $\slogm$ below
the observed value, leaving room for other sources of scatter from
more physical effects.

Throughout this paper, we assume a flat-\lcdm\ cosmology with
$\om=0.3$, $\s8=0.8$, and $h=0.7$.  We will use redshift as our time
unit quite often, especially in our model parameterizations, but will
show plots as expansion factor $a$, which is a more natural time unit
for the growth of galaxies.

\begin{figure*}
\includegraphics[width=7in ]{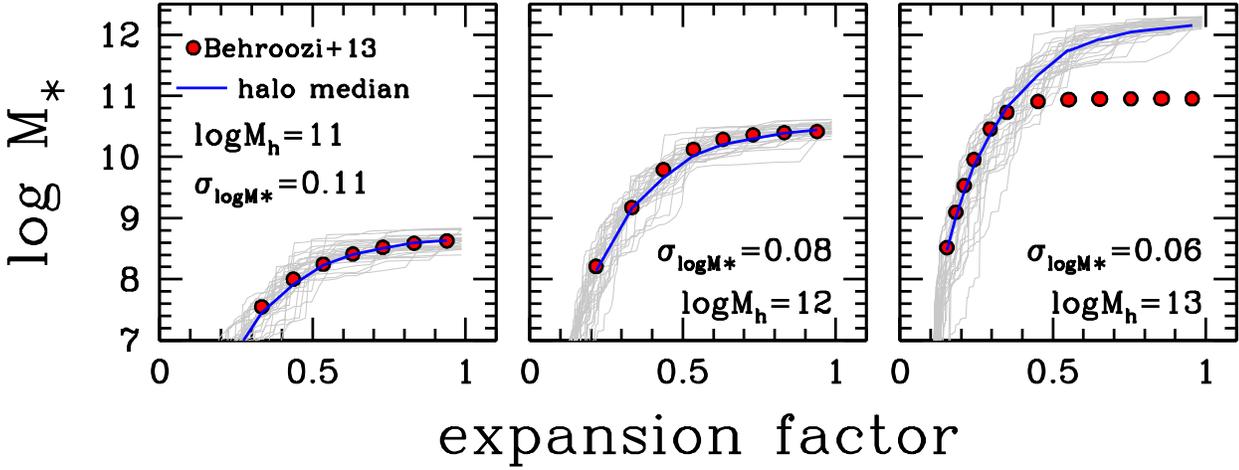}
\vspace{-11.cm}
\caption{ \label{galpath_hmass} The stellar mass growth of halos of
  various masses. In each panel, the solid blue curve is the median
  value from a series of halo merger trees. This curve is fit to the
  results of \citet{behroozi_etal:13}, which are shown with the red
  circles. The evolutionary tracks of a subsample of halos are shown
  in the thin gray curves. For each halo, the same $\fcon(z)$ function
  is applied, thus the differences in stellar mass are all driven by
  the differences in halo mass growth. In each panel, the amplitude of
  $\fcon(z)$ is varied but the redshift-dependence is the same. This
  figure shows three things: (1) a single function of redshift con
  describe $\fcon$ for halos $\le 10^{12}$ \msol, as well as the
  high-redshift growth in massive halos (2) that $10^{13}$ \msol\
  halos undergo a rapid transition in their star formation efficiency
  at $z\sim 2$. This is a restatement of the previous results found in
  abundance matching studies And (3), different halo formation
  histories impart a scatter of $\la 0.1$ dex in $\log\mgal$ for a
  generic $\fcon(z)$ function. }
\end{figure*}

\begin{figure*}
\includegraphics[width=7in ]{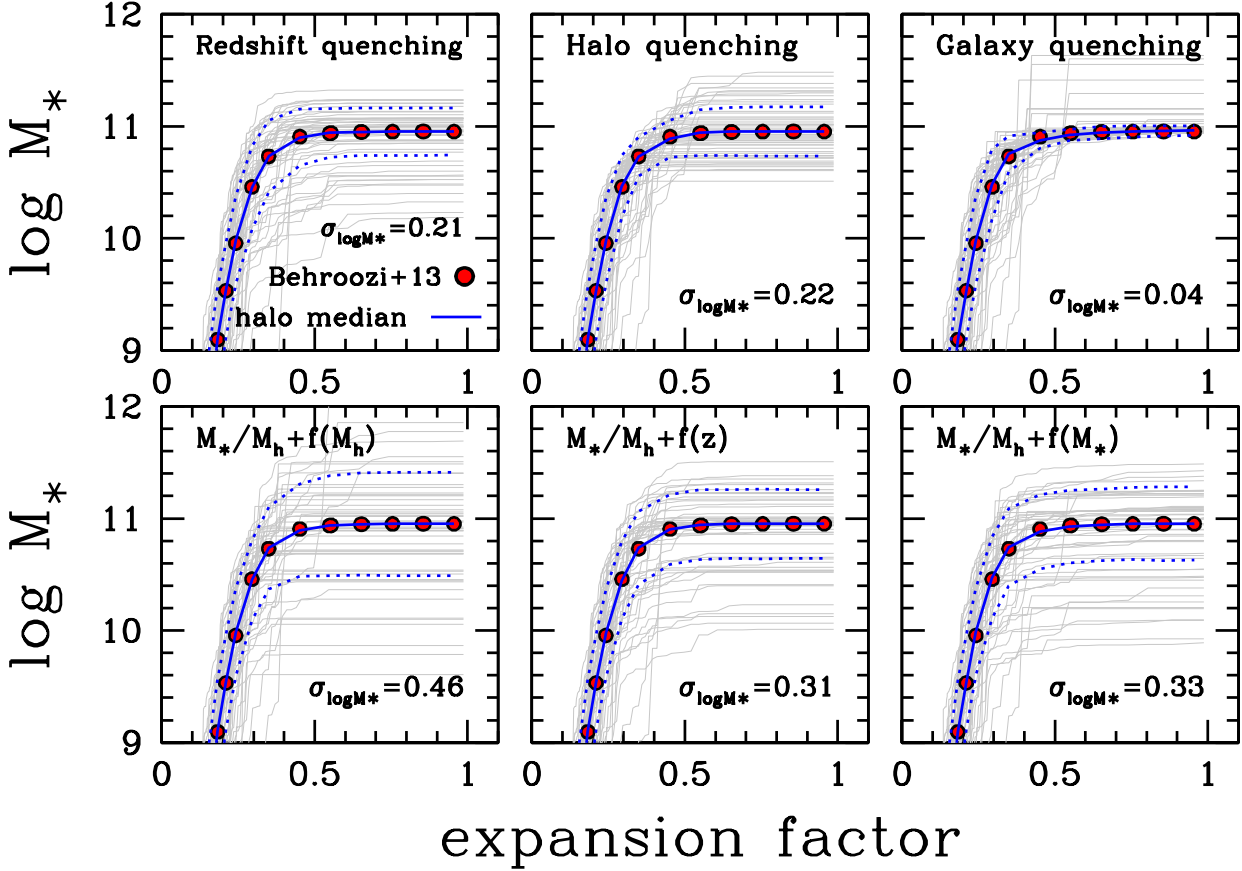}
\vspace{-5.5cm}
\caption{ \label{galpath6} {\it Top Row:} Best fit models for
  redshift quenching, halo quenching, and galaxy quenching. The solid
  blue curves are the median of the set of halo merger trees, which
  are fit to the data of \citet{behroozi_etal:13}, shown with the red
  circles. A sample of individual halos are shown with the gray
  curves. The value of the scatter, $\slogm$, is shown for each model
  in the panel. The blue dotted curves show the 68\% range of $\mgal$
  around the median. {\it Bottom Row:} Same as the top row, but now
  for the three different models where the threshold for quenching is
  based on a critical $\mgal/\mhalo$ ratio, and after that point
  $\fq$ is parameterized by $z$, $\mhalo$, and $\mgal$,
  respectively. }
\end{figure*}

\section{Models}

\subsection{Parameterizing Star Formation Efficiency}

We define the baryon conversion efficiency as

\begin{equation}
\fcon\equiv SFR\times \left[\frac{\omb}{\om} \dot{\mhalo}\right]^{-1},
\end{equation}

\noindent where $SFR$ is star formation rate, $\omb/\om$ is the
universal baryon fraction (which we assume to be $0.045/0.3=0.15$),
and $\dot{\mhalo}$ is the instantaneous growth rate of the
halo. Inspection of the abundance matching results of
B13 and \cite{moster_etal:13} shows that for
low-mass halos, and in the absence of any external quenching
mechanism, $\fcon$ can approximately be parameterized as a
function that depends only on redshift:

\begin{equation}
\fcon(z) =  \left\{ \begin{array}{ll}
		f_0({\mhalo}_0)\left(\frac{1+z}{1+z_0}\right)^{\gamma_1}
                & {\rm if\ \ } z>z_0 \\ 
 & \\
		f_0({\mhalo}_0)\left(\frac{1+z}{1+z_0}\right)^{\gamma_2}
                & {\rm if\ \ } z\le z_0\\
		\end{array}
	\right.
\end{equation}

\noindent where $\gamma=-3$ at $z>z_0$ and $\gamma=0$ and $z\le z_0$,
$f_0({\mhalo}_0)$ is an overall amplitude that depends on $z=0$ halo mass,
and $z_0=1$. The total stellar mass at any redshift $z$ is

\begin{equation}
\label{e.intmgal}
\mgal(z) = \int_0^{t(z)} SFR(t)\,dt = \int_\infty^z \fcon(z^\prime) f_b \dot{\mhalo} \frac{dt}{dz^\prime} dz^\prime.
\end{equation}

\noindent To calculate the stellar mass growth for an individual halo,
we use numerical halo merger trees (described in the next
subsection). The trees calculate the mass of the halo at discrete time
intervals, thus rather than implement equation (\ref{e.intmgal})
directly we use discrete summation over the timesteps, assuming that
the integrand is a constant in time over each interval.

\begin{equation}
\label{e.summgal}
\mgal(z_i) = \sum_i \fcon(z_i)f_b\Delta M_{h,i}.
\end{equation}

\subsection{Parameterizing Star Formation Quenching}
\label{s.model_quench}

We parameterize the quenching of star formation by the quantity $\fq$,
such that $SFR(z) \propto \fcon(z)\times\fq(z)$. Figure
\ref{galpath_hmass} shows the time evolution of $\mgal(z)$ in
present-day $10^{13}$ \msol\ halos, as derived by
B13.  From these results, it is clear that
quenching in these halos must happen over a short timescale, as
stellar mass growth at $z<1$ is almost negligible. Thus we
parameterize $\fq$ as an exponential function with free parameters
governing the onset of quenching and its rapidity. We consider 6
different models for parameterizing the time evolution of $\fq$. The
forms implemented are all listed in Table \ref{table1}.

\begin{itemize}

\item {\it Redshift quenching}: quenching begins at $z<z_{\rm
    crit}$. Redshift quenching is a fairly ad-hoc model, although one
  can conceive of redshift-dependent quantities that may impact star
  formation. The results of this model can be thought of as applying
  the $\fcon$ results of B13 (or \citealt{moster_etal:13}) and applying them
  to individual halos.

\item {\it Halo quenching}: quenching begins when $\mhalo(z)>\mhcrit$. The
  idea of a critical halo mass beyond which galaxy formation is
  curtailed is driven by numerical simulations that demonstrate that
  gas accretion onto halos undergoes a rapid transition at $\sim
  10^{12}$ \msol\ (\citealt{keres_etal:05, keres_etal:09,
    dekel_birnboim:06}). Below $\mhcrit$, gas is accreted cold and is
  deposited directly onto the central galaxy. Above this threshold,
  gas is shock heated to high temperature and gas cooling is
  significantly attenuated. Interpreting this threshold as a threshold
  for quenching star formation leads to a natural explanation of
  galaxy bimodality (\citealt{cattaneo_etal:06}).

\item {\it Galaxy quenching}: quenching begins at
  $\mgal(z)>\mgcrit$. The bimodality of galaxies can be seen most
  clearly in their stellar mass (e.g., \citealt{kauffmann_etal:03}),
  with a clear break in their $z=0$ properties at $\mgal\approx
  10^{10.3}$ \msol. Such a scenario could be induced by instabilities
  in disk galaxies that occur after the disk becomes too massive. In
  the semi-analytic model of \cite{bower_etal:06}, disk instabilities
  are the primary feeding mechanism for the central black hole, and
  thus the source of galaxy quenching. This model is described as
  `secular evolution' in the \cite{hopkins_etal:08b}.

\item {\it Ratio quenching}: quenching begins when the ratio of the
  stellar mass to halo mass reaches a peak value,
  $\ratio=\ratiocrit$. This threshold was proposed by
  \cite{leauthaud_etal:12_shmr} to explain the apparent lack of
  evolution of peak value of $\ratio$ from $z=1\rightarrow 0$. The
  stellar to halo mass ratio is not monotonically rising, however, so
  the ratio itself is only used to determine the redshift of the onset
  of quenching, and we employ a secondary parameter to parameterize
  $\fq$. Table \ref{table1} lists the three different implementations
  of $\ratio$-quenching, where $\fq$ is parameterized by $z$,
  $\mhalo$, and $\mgal$. Where the critical value of each quantity is
  determined by the redshift at which $\ratio=\ratiocrit$.

\end{itemize}

Each model has four free parameters: $f_0$, $\sigma_{(x)}$,
$\alpha_{(x)}$, and the critical threshold in parameter $x$ that
induces the quenching. The two remaining parameters in $\fcon$ are
fixed to the best-fit values obtained from the lower-mass halos:
$z_0=1$ and $\gamma=-3.0$. We explore the posterior distributions of
the free parameters using Markov Chain Monte Carlo (MCMC), using
$\chi^2$ of each model with respect to the B13 measurement of
$\mgal(z)$ in $10^{13}$ \msol\ halos to estimate the likelihood of
each model. Due to asymmetries in some of the distributions of
$\mgal(z)$ induced over the set of halo merger trees, we use the
median value of $\mgal(z)$ rather than the mean, as well as the 68\%
range of values to estimate the scatter $\slogm$. The results of the
MCMC chains are shown in Table \ref{table2}. These will be discussed
in the following section.

The models described above all assume that the quenching threshold, in
each quantity, is a constant in time. There are two straightforward
extensions of these models that we will discuss further in the draft.

\begin{itemize}

\item {\it Time-varying treshold:} Results of theoretical models of
  critical halo mass indicate that there is little, if any, variation
  of the transitional mass scale between cold and hot accretion, but
  other effects may come into play, such as redshift evolution of
  metallicity. After presenting results of constant quenching
  thresholds in \S \ref{s.results_constant}, we will incorporate a
  time-varying quenching barrier (for models that allow such freedom,
  which therefore excludes redshift quenching). We implement a
  straightforward linear dependence of the critical quantity on
  expansion factor $X_{\rm
    crit} = X_{\rm crit,0} + (a-0.3)\times \beta$. Where $X$
  represents $\mhalo$, $\mgal$, or $\ratio$. 

\item {\it Stochastic quenching:} We allow the the critical threshold
  to vary in a stochastic manner from halo to halo using a random
  Gaussian deviate for halos $i$, i.e., $X_{{\rm crit}, i} = X_{\rm
    crit} + G(\sigma_{\rm stoch})$ where $G$ is a Gaussian with zero
  mean and width $\sigma_{\rm stoch}$. For halo and galaxy quenching
  $X\equiv \log \mhcrit$ and $X\equiv \log\mgcrit$, and for redshift and
  ratio quenching, $X$ is linear in $z$ and $\ratio$. 

\end{itemize}


\begin{table*}
\centering
\begin{tabular}{@{}lll@{}}
\hline
Model & Form & Notes \\
\hline

\vspace{0.1in}
Redshift Quenching & $\fq(z) =
\exp\left(\frac{z-\zcrit}{\sigma_z}\right)^{\alpha_z}$ & \\
\vspace{0.1in}
Halo Quenching &  $\fq(\mhalo) =
\exp\left(\frac{\log\mhcrit-\log\mhalo}{\sigma_h}\right)^{\alpha_h}$ &
\\
\vspace{0.1in}
Galaxy Quenching & $\fq(\mgal) =
\exp\left(\frac{\log\mgcrit-\log\mgal}{\sigma_g}\right)^{\alpha_g}$ &
\\
\vspace{0.1in}
Ratio Quenching $+f_Q(z)$ & $\fq(z) =
\exp\left(\frac{z-\zcrit}{\sigma_z}\right)^{\alpha_z}$ &
$\zcrit = z(\ratio=\ratiocrit)$ \\
\vspace{0.1in}
Ratio Quenching $+f_Q(\mhalo)$ & $\fq(\mhalo) =
\exp\left(\frac{\log\mhcrit-\log\mhalo}{\sigma_h}\right)^{\alpha_h}$ &
$\mhcrit = \mhalo(\zcrit)$
\\
\vspace{0.1in}
Ratio Quenching $+f_Q(\mgal)$& $\fq(\mgal) =
\exp\left(\frac{\log\mgcrit-\log\mgal}{\sigma_g}\right)^{\alpha_g}$ &
$\mgcrit = \mgal(\zcrit)$
\\

\end{tabular}
\caption{Parameterization of the quenching function $\fq$. }
\label{table1}
\end{table*}

\subsection{Halo Merger Trees}

We use merger trees created by the publicly available code of
\cite{neistein_dekel:08}. This algorithm is based off the extended
Press-Schechter formalism but calibrated to match the results of
numerical simulations. Although the code is able to create full trees
for each final halo, the only quantity we are interested in is the
time evolution of the mass of the main progenitor, for which the code
is especially accurate.

\begin{figure*}
\includegraphics[width=7in ]{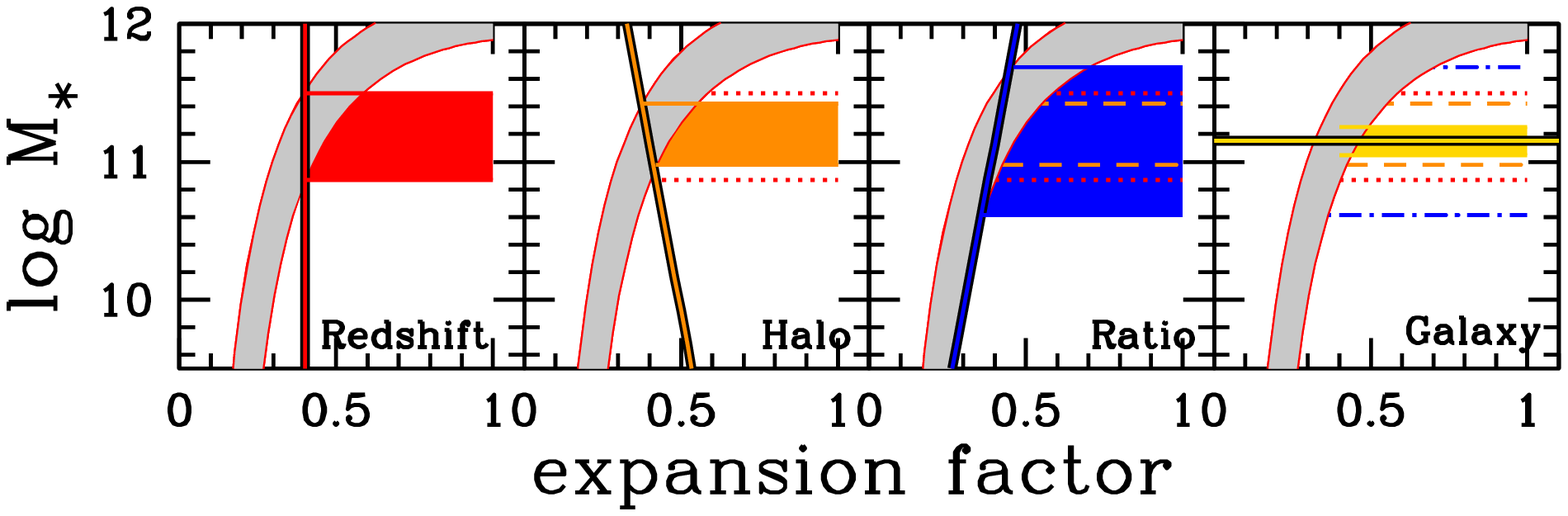}
\vspace{-12cm}
\caption{ \label{pedagogical} Pedagogical explanation for the
  different scatters induced by different quenching mechanisms. The
  gray shaded region in each panel represents the growth of galaxies
  in the absence of any quenching (c.f. Figure
  \ref{galpath_hmass}). Each physical property produces a threshold
  barrier with a different slope in the above diagram. The scatter is,
  to a reasonable approximation, the range on the $y$-axis where the
  barrier intersects the upper and lower bounds of the standard galaxy
  growth curves. Redshift scatter is thus a vertical barrier, and the
  scatter subsequent to quenching is the red shaded region.  Halo
  quenching yields a barrier with a negative slope, shrinking the
  scatter. $\ratio$ quenching creates a threshold with steep positive
  slope, expanding the range of the post-quenching galaxy
  masses. Galaxy quenching simply stops growth at a constant
  horizontal barrier, yielding arbitrarily small scatter. More
  discussion on the slope of these barriers can be found in the text.
}
\end{figure*}

\begin{figure}
\includegraphics[width=3.5in ]{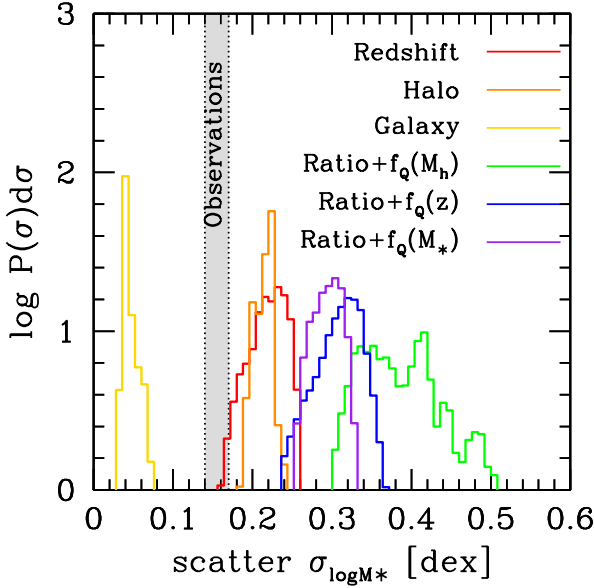}
\vspace{-1cm}
\caption{ \label{scatter} The posterior distributions of $\slogm$ for
  each quenching model after being fit to the \citet{behroozi_etal:13}
  stellar mass growth curves. The observations are taken from
  clustering measurements of BOSS galaxies in
  \citet{tinker_etal:16_boss}.}
\end{figure}

\begin{table*}
\centering
\begin{tabular}{@{}lllll@{}}
\hline
Model & $f_0$ & $X_{\rm crit}$ & $\sigma_{(x)}$ & $\alpha_{(x)}$ \\
\hline

Redshift & $ 0.34 \pm 0.02$ & $ 3.87 \pm 0.72$ & $ 1.85 \pm 0.69$ & $ 3.42 \pm 1.00$ \\ 
Halo & $ 0.34 \pm 0.02$ & $ 11.66 \pm 0.23$ & $ 0.58 \pm 0.22$ & $ 3.63 \pm 0.85$ \\ 
Galaxy & $ 0.33 \pm 0.02$ & $ 10.11 \pm 0.27$ & $ 0.52 \pm 0.22$ & $ 3.60 \pm 0.94$ \\ 
Ratio+$f(z)$ &$ 0.34 \pm 0.02$ & $ 0.019 \pm 0.006$ & $ 1.26 \pm 0.47$ & $ 3.03 \pm 1.08$ \\ 
Ratio+$f(\mhalo)$ &$ 0.33 \pm 0.02$ & $ 0.024 \pm 0.006$ & $ 0.33 \pm 0.16$ & $ 3.03 \pm 1.12$ \\ 
Ratio+$f(\mgal)$ & $ 0.33 \pm 0.02$ & $ 0.027 \pm 0.006$ & $ 0.37 \pm 0.19$ & $ 3.36 \pm 1.02$ \\ 

\hline

Redshift & $ 0.34 $ & $ 3.14 $ & $ 1.10 $ & $ 2.14 $ \\ 
Halo & $ 0.33 $ & $ 11.40 $ & $ 0.84 $ & $ 4.81 $ \\ 
Galaxy & $ 0.33 $ & $ 9.91 $ & $ 0.73 $ & $ 4.95 $ \\ 
Ratio+$f(z)$ &$ 0.34 $ & $ 0.017 $ & $ 0.51 $ & $ 4.07 $ \\ 
Ratio+$f(\mhalo)$ &$ 0.34 $ & $ 0.014 $ & $ 1.71 $ & $ 3.36 $ \\ 
Ratio+$f(\mgal)$ & $ 0.34 $ & $ 0.021 $ & $ 0.60 $ & $ 4.79 $ \\ 

\end{tabular}
\caption{Constraints on the quenching models. The top 6 lines show the
mean and variance of each parameter in the model. The bottom 6 lines
show the values of the best-fit model. Correlations between parameters
and asymmetries in the posterior distributions of each parameter yield
best-fit parameters that deviate from the mean values from the chains.}
\label{table2}
\end{table*}

\section{Results}

\subsection{Scatter in Intrinsic Physical Processes of Star Formation}

Figure \ref{galpath_hmass} shows the results of applying equation
(\ref{e.summgal}) to the numerical halo merger trees of various $z=0$
masses. These results do not include any quenching. At each halo mass,
we fit for the amplitude of $\fcon$, $f_0$, by comparing the median of
the simulated halos to the B13 data. For each halo mass considered,
the best-fit value of $f_0$ is $0.05$, $0.27$, and $1.40$,
respectively, going from low mass to high mass\footnote{Note that
  $f_0>1$ for these halos does {\it not} imply that they are
  converting more than 100\% of their baryons into stars, as $\fcon$
  goes as $(1+z)^{-3}$ at the redshifts where this model is applied to
  massive halos.}. For $\mhalo\le 10^{12}$ a simple function for
$\fcon$ fits the results for various halo masses with only one
tunable parameter. For massive halos, the results are quite different.
As there is no quenching implemented in this figure, the models for
$10^{13}$ \msol\ halos are only fit to data at $a<0.3$ $(z\ga
2$)\footnote{Fitting E.g. (4) to the data at all times ledas to a
  best-fit model being a poor description of the data at all
  redshifts: the attenuation in stellar growth is too slow. The fit
  here is performed for pedagogical purposes only.}. The
rapid attenuation of star formation at $a\sim 0.3$ ($z\sim 2$) is
clear in these models. At $a>0.3$, it is clear that the `universal'
$\fcon$ function no longer applies to massive halos, and some form of
process most occur to nearly extinguish the growth of the galaxies
that form within them.

The thin gray lines show a subsample of the galaxy growth curves for
individual halos. Although all halos in each panel are the same $z=0$
mass, the individual trajectories can differ substantially, especially
at high redshift. Early forming halos accrete most of their mass (and
thus most of their baryons) at redshifts when $\fcon$ is lower, thus
they will have lower stellar masses at $z=0$ than late-forming
halos. At late times ($z<z_0$), $\fcon$ becomes independent of
redshift, so any scatter in $z=0$ galaxy masses is fixed at that
time. The values of $\slogm$ monotonically decrease at halo mass
increases. This is because larger halos grow at faster rates than
smaller halos, thus their galaxies grow more from $z=z_0\rightarrow
0$. The larger the fraction of stellar mass is created over this time
frame, the more the scatter induced from growth during $z>z_0$. 

The values of $\slogm$ in Figure \ref{galpath_hmass} are all
comfortably below the observed value of 0.16 dex. However, the results
of this figure put limits on all other sources of scatter (including,
for example, having a non-universal $\fcon$ that varies from halo to
halo in a systematic way that does not correlate with halo
formation history). 

\begin{figure*}
\includegraphics[width=6in ]{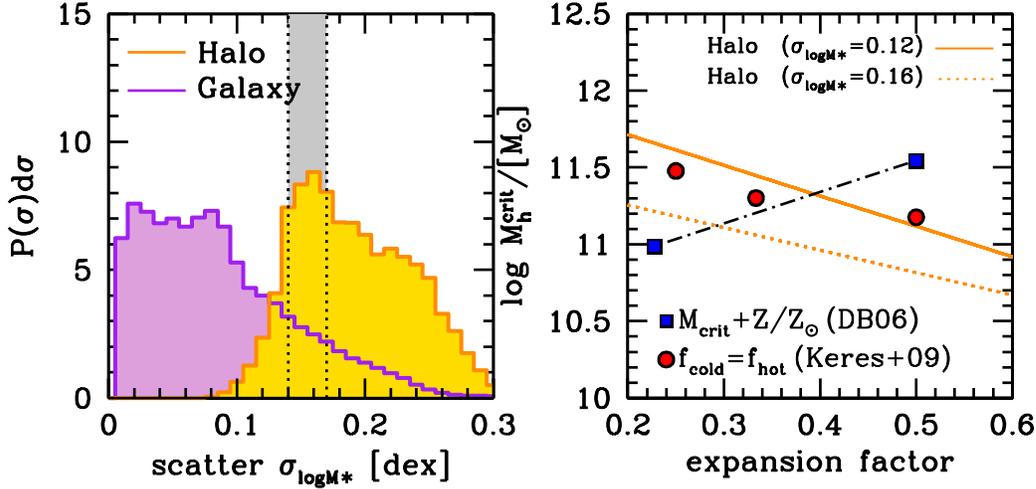}
\vspace{-7cm}
\caption{ \label{threshold} {\it Left panel}: The posterior
  distribution of $\slogm$ for galaxy and halo quenching models in
  which the quenching threshold varies linearly with expansion
  factor. This freedom brings the halo quenching model into agreement
  with observational constraints on $\slogm$, yielding a 95\% lower
  limit of 0.12 dex. The extra freedom also increases the scatter
  obtained in galaxy quenching, allowing value much higher than 0.16
  dex. {\it Right Panel}: How the halo threshold changes with $a$ in
  the various models. The two orange curves show the halo quenching
  threshold for $\slogm=0.16$ dex and for the lower limit of 0.12 dex.
  The red circles show the halo masses at which cold and hot gas
  accretion are equal in the simulations of \citet{keres_etal:09}. The
  blue squares connected by the dash-dot line show how $\mhcrit$ would
  vary with time due to the increase in metallcity in massive galaxies
  over cosmic time (inferred from \citealt{dekel_birnboim:06}; see
  text for further details and caveats.)}
\end{figure*}

\subsection{Results with a Constant Quenching Threshold}
\label{s.results_constant}

Figure \ref{galpath6} shows the results of the best-fit parameter sets
from the all models in \ref{table1}. The panels show $\mgal(z)$ for
the median of all halos, as well as a subsample of individual
halos. In each panel, the scatter induced in each model is also
listed; both redshift and halo quenching induce a scatter larger than
the observed value of $\slogm=0.16$. Galaxy quenching, unsurprisingly,
creates a minimal scatter. For redshift quenching, $\fq(z)$ is the
same for all halos. The product of $\fcon(z)\times\fq(z)$ yields the
mean baryonic conversion efficiency of $10^{13}$ \msol\ halos found in
B13 (c.f. their Figure 11). Halo quenching produces results very
similar to redshift quenching, but the onset of quenching now varies
from halo-to-halo. The similarity in the two models is due to the fact
that $\log M\propto -z$ (\citealt{wechsler_etal:02}), and so the mean
$\fq(\mhalo)$ is very similar to $\fq(z)$. The added
variation in the time of quenching does not add to the $z=0$ variance
in $\mgal$, as we will discuss with Figure \ref{pedagogical}.

The bottom row of Figure \ref{galpath6} shows the results from the
second three models in \ref{table1}, where the quenching threshold is
defined by $\ratio$, but $\fq$ itself is parameterized by $z$,
$\mhalo$, and $\mgal$. For these models, $\slogm$ is significantly
larger than the first three. The different behavior of all six models
can be understood in the schematic outlines in Figure
\ref{pedagogical}. The gray shaded region shows the $\pm 1\sigma$
range of evolutionary tracks of galaxies within $10^{13}$ \msol\ halos
in the absence of any quenching (i.e., the right-hand panel of Figure
\ref{galpath_hmass}). The thick colored lines indicate the `quenching
barrier' implied by each model in Table \ref{table1}. Redshift
quenching imposes a simple vertical barrier at fixed time. The range
of post-quenching galaxy masses is indicated by the points on the
$y$-axis where this barrier intersects the upper and lower bounds of
the galaxy evolutionary tracks. For halo quenching, galaxies evolving
on the $+1\sigma$ track form in halos that grow rapidly at high
redshift. Thus, these halos hit the threshold value of $\mhalo$ at
earlier times than those galaxies evolving on the $-1\sigma$ track in
slower-forming halos. Due to the tilted nature of the halo mass
barrier, the range of galaxy mass is smaller than for redshift
quenching. In detail, the value of $\slogm$ does vary with the exact
values of the free parameters in the models, thus it is possible to
produce a redshift quenching model with smaller $\slogm$ than a halo
quenching model. However, as we will show in Figure \ref{scatter}, the
median $\slogm$ for halo quenching models is somewhat smaller than
that for redshift quenching.

For ratio quenching, the trend of the barrier with redshift is
opposite from halo quenching. At high redshift, halos grow faster than
galaxies because $\fcon \propto (1+z)^{-3}$. Thus, rapid-forming halos
take longer to create enough stellar mass to meet the $\ratiocrit$
treshold. Halos that form later are accreting more of their mass at a
time when $\fcon$ is higher and baryonic conversion is thus more
efficient. Therefore these halos match $\ratiocrit$ earlier. The net
effect is that the 1$\sigma$ range in post-quenching $\mgal$ is spread
out significantly relative to the other models. The rightmost panel of
Figure \ref{pedagogical} shows galaxy quenching barrier as a
horizontal line at fixed $\mgal$. Thus, this model yields minimal
scatter. The small scatter induced in the models is due to variations
in halo growth over the period of time after quenching begins but
before quenching is complete, or galaxies that grow significantly in
the timestep in which they cross the barrier. In equation
\ref{e.summgal}, quenching only begins on the timestep after the halo
crosses the threshold. This is a choice in the implementation of the
model, but would physically represent a scenario in which a halo
crosses the threshold during a major starbursting event (such as a
merger), and the quenching does not begin until after that burst of
star formation declines.

Figure \ref{scatter} shows the posterior distributions of $\slogm$ for
each model, marginalizing over the other parameters. The measurements
of $\slogm$ for $10^{13}$ \msol\ halos from BOSS galaxies
(\citealt{tinker_etal:16_boss}) are shown with the gray shaded
area. These measurements are at $z\sim 0.5$, but the dispersion in
$\mgal$ in the models is essentially constant after quenching
completes at $z\sim 1$. The features and asymmetries in the posterior
distributions are real and not the result of noise or lack of
convergence in the MCMC chains. The results here reflect those seen in
the best-fit models of Figure \ref{galpath6}; halo quenching and
redshift quenching models yield results near $\slogm\sim 0.22$, while
the different ratio quenching models yield values of $\slogm$ that are
much too high, ranging up to $\ga 0.5$ for ratio+$\fq(\mhalo)$. Galaxy
quenching yields $\slogm \sim 0.04$.

\subsection{Increasing Freedom in the Models}

As discussed in \S \ref{s.model_quench}, models other than redshift
quenching can be relaxed to allow the quenching threshold to vary in
time. We have repeated the analysis for 5 of the 6 models that allow
this freedom. The results for ratio quenching do not notably change,
thus we will focus on halo and galaxy quenching from this point
forward in this section.

The left panel in Figure \ref{threshold} shows the posterior
distribution of $\slogm$ with this added freedom. The range of
$\slogm$ values is significantly expanded for both models, and both
now overlap with the observed value. The right panel in Figure
\ref{threshold} shows examples of $\mhcrit(z)$ that yield good fits to
$\mgal(z)$ for two $\slogm$ values. For halo quenching, models that
reduce scatter have a tilted threshold such that $\mhcrit$ decreases
with time. In the constant-$\mhcrit$ model, late-forming halos will
generally form galaxies above the mean. Tilting the threshold such
that it is higher at higher redshift allows early-forming halos to
convert a higher threshold of their baryons into stars before the
onset of quenching, reducing the correlation between formation history
and $z=0$ $\mgal$. As shown in the Figure, the slope of the barrier is
independent of $\slogm$ for values that are at and below the observed
value, but the amplitude increases as $\slogm$ decreases. This
enforces a lower limit on the value of $\slogm$ achievable with halo
quenching; quenching must by nearly complete by $z\sim 1.5$, when most
halos at $\sim 10^{12}$ \msol.

The theoretical expectation for how $\mhcrit$ should vary---if at
all---with redshift is not clear. Figure \ref{threshold} shows results
from two of the canonical studies. The filled circles show the
transition halo mass scale between cold mode and hot mode accretion in
the hydrodynamical cosmological simulations of \cite{keres_etal:09},
defined here as the halo mass scale where the fraction of gas being
accreted in the hot and cold modes are equal. In this model, there is
a slight decrease in $\mhcrit$ with time that is in reasonable
agreement with the data. However, one major simplification in these
results is that there is no metal enrichment in the gas as the
simulation evolves. \cite{mannucci_etal:09} find that the metallicity
for massive galaxies increases by a factor of $\ga 5$ from $z=3$ to
$z=0.7$ (parameterized as $\zz$. \cite{dekel_birnboim:06}, using
one-dimensional hydro simulations, find a strong dependence of
$\mhcrit$ with metallicity. Combining these results with the
observational measurements of $\zz$, $\log\mhcrit$ should increase by
nearly half a decade over the timespan of active quenching, shown by
the filled squares in the Figure. This calculation makes the
assumption that gas metallicity only a function of time, independent
of both the halo and galaxy formation history. Although this is likely
to be an oversimplification in detail, the main point is that any
increase of $\mhcrit$ with time has the net effect of increasing
$\slogm$, and $\zz$ increases with time.

Scatter can be significantly increased in the galaxy quenching model
by a time-varying $\mgcrit$. Since a `horizontal' quenching barrier
produces no scatter, tilting this barrier in either direction will
have the net effect of increasing this scatter.

\subsection{Stochasticity in the Models}

A more physical model may include some stochasticity in
quenching---i.e., that the value of $\mhcrit$ may vary randomly from
halo to halo. To implement stochasticity, we choose a random Gaussian
variable with zero mean and dispersion $\sigstoch$, which for the halo
quenching model is in units of dex (i.e.,
$\log\mhcrit$). Stochasticity will only increase $\slogm$, but it is
important to ask how much stochasticity is allowed by available data. 

\begin{figure}
\includegraphics[width=3.4in ]{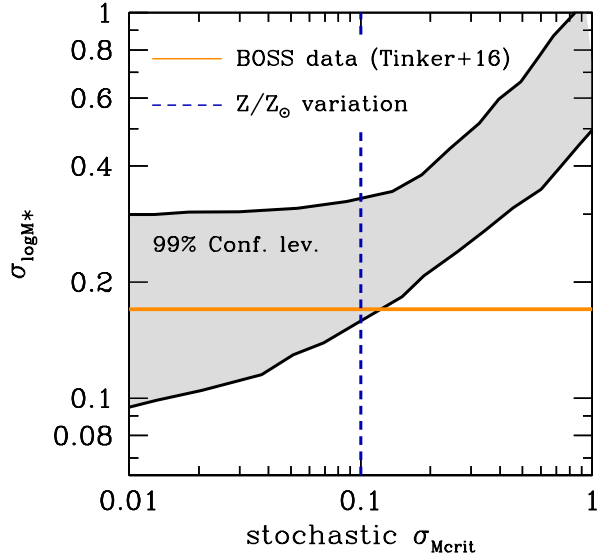}
\vspace{-1cm}
\caption{ \label{stochasticity} The effect on $\slogm$ when including
  stochasticity in the halo quenching treshold. Stochasticity only has
  the effect of increasing the induced scatter. The shaded region
  shows the 99\% confidence region for this model. The horizontal line
  indicates the measured value of $\slogm$ from BOSS data. The dashed
  vertical line indicates the variation in $\log\mhcrit$ induced by
  the scatter in metallicity, assuming that this scatter is
  uncorrelated with halo formation history. }
\end{figure}

Figure \ref{stochasticity} shows the range of models in the
$\slogm$-$\sigstoch$ plane for the halo quenching model. In this
analysis, we adopt of flat prior on $\log\sigstoch$. This model
includes a time-varying quenching threshold. The lower envelope
delineated by the chain indicates that, in order to be consistent with
the data, any stochasticity in $\log\mhcrit$ must be lower than 0.1
dex.

Calling once again on the dependence of $\mhcrit$ on metallicity found
in \cite{dekel_birnboim:06}, random variations of $\zz$ between halos
may induce such stochasticity in quenching. The scatter in $\zz$
around $10^{11}$ \msol\ galaxies is 0.2 dex \citep{gallazzi_etal:05},
which translates into a variation of 0.1 dex in $\log\mhcrit$
(\citealt{dekel_birnboim:06}, their Figure 4). Thus, if variations of
$\zz$ between galaxies of the same mass is random, this stochasticity
would not allow any other contribution to $\slogm$ and still be in
agreement with the data. This is a strong assumption, but if
metallicity is correlated with halo formation history it may be
detectable through the effect of formation history on clustering.

\section{Summary and Discussion}

We have implemented a simple model to explore how different models for
quenching star formation in galaxies can impact the scatter of stellar
mass at fixed halo mass, $\slogm$, for which we have excellent
constraints from the clustering of massive galaxies. We test models in
which quenching begins at some critical redshift, $\mhalo$, $\mgal$,
or $\mgal/\mhalo$ ratio. We find:

\begin{itemize}
\item Under the assumption that the quenching threshold is constant
  with time, only galaxy quenching is consistent with the measurements
  of $\slogm$.
\item The scatter imparted by halo and redshift quenching is somewhat
  larger than observations, while the scatter yielded by ratio
  quenching is nearly double that observed.
\item To decrease the scatter induced by halo quenching, $\mhcrit$
  must decrease with time. This goes in the opposite direction
  implied by the growth of metallicity in galaxies.
\item There is little to no room for any stochasticity in $\mhcrit$
  from halo-to-halo. The observed scatter in $\zz$, if uncorrelated
  with halo formation history, would raise $\slogm$ above the observed
  values.
\item Decreasing the scatter in each model would require strong
  correlations between galaxy properties, such as metallicity or
  mean stellar age, and halo formation history.
\end{itemize}

Although galaxy quenching yields, by far, the lowest values of
$\slogm$, there are some obvious questions that arise from this
model. Observations of central galaxies within halos show that the
quenched fraction varies smoothly with increasing galaxy mass, and not
consistent with a threshold value
(\citealt{weinmann_etal:06a}). Stochasticity in the $\mgcrit$ threshold
may alleviate this tension, as well as some correlation of another
galaxy property with the quenching threshold. Applying this model to
the full galaxy population, rather than just massive galaxies, can
resolve this question. This will be pursued in a future paper.

There are many simplifications and assumptions that are used to
construct the models described above, but many---if not most---create
testable predictions that may be constrained by existing data. The
current spectroscopic sample of massive galaxies presently contains
upwards of $\sim 2$ million galaxies and reaches over 7 Gyr into the
cosmic past through the combination of SDSS, BOSS, and now eBOSS data
(for which the first clustering measurements have been published;
\citealt{zhai_etal:16}). Although the quality of many of these spectra
make detailed stellar population analysis untenable on a per-object
basis, the clustering of these galaxies contains a wealth of
information, well beyond the value of $\slogm$ used here. At fixed
$\mgal$, the dependence of any galaxy property on halo formation
history will show up in the clustering of those galaxies. This halo
assembly bias has been shown conclusively in numerical simulations,
and has recently been detected in cluster-sized dark matter halos
observationally (\citealt{miyatake_etal:16,
  more_etal:16}). \cite{saito_etal:16} used two-point clustering to
demonstrate that $z\sim 0.5$ BOSS galaxies are consistent with a model
in which the colors of massive galaxies are correlated with halo age.

For the fiducial implementations of galaxy and halo quenching, in
which the threshold is constant in time, assembly bias is a natural
consequence. Early-forming halos will have older stellar populations
at fixed $\mgal$, which could impart an assembly bias signal based on
galaxy color, luminosity, and metallicity at fixed mass.  Redshift and
ratio quenching, on the other hand, yield little correlation between
mean stellar age and the halo formation time. Different
implementations of the halo quenching model yield different assembly
bias signals as well.  Halo quenching models with a time-varying
threshold tend to reduce the amount of assembly bias in the models
because they delay quenching in early-forming halo and accelerate it
in late-forming halos. Further investigation, both through the
clustering of massive galaxies and by using stellar population
synthesis models to calculate the observable properties of galaxies
with various mean stellar ages, will be fruitful in further
differentiating models or constraining a the parameter space of a
specific model.

We have assumed that other physical mechanisms that effect star
formation within dark matter halos would add to the scatter induced by
variations in halo formation history. We have specifically focused on
metallicity as a probable source of such scatter within the halo
quenching model, either in the form of stochastic variations of the
quenching threshold or a redshift dependence that would widen the
scatter over the fiducial model. It is always possible that these
physical mechanisms correlate with halo formation history in a way to
reduce the scatter in stellar mass. For example, if metallicity
correlated with halo formation history such that early-forming halos
have higher metallicity that later-forming halos, the dependence of
$\mhcrit$ on $\zz$ would help reduce scatter by allowing early-forming
halos to convert more of their baryons into stars than they would in
the constant threshold model. Such a model would create an assembly
bias signal on the metallicity of massive galaxies. 

Another assumption we have made in the construction of these models is
that $\fcon$ is a universal function that only depends on
$z$. Reducing the scatter in the pre-quenching phase of evolution
would also reduce post-quenching scatter. Given the large variation in
halo mass at $z=3$ for present day $10^{13}$ \msol\ halos---roughly a
factor of five---it is difficult to construct a model that creates
minimal scatter in $\mgal$ within these halos at high redshift that
isn't highly ad hoc. For $z=0$ halos below the quenching threshold,
there still exists a scatter in stellar mass that is larger than that
shown in Figure \ref{galpath_hmass}. A model in which $\fcon$ depends
on $z$ and some second parameter, such as $\mhalo(z)$ or $\mgal(z)$,
may shrink the scatter at $z\sim 2$, but it is not clear that such a
model would yield small scatter at $z=0$, as well as reproduce the
measurements of SFR$(z)$ for galaxies of various masses, which is well
fit by Equation (\ref{e.intmgal}) (B13, \citealt{moster_etal:13}). 

Any strong conclusions made here depend on strong
assumptions. However, this work represents a proof-of-concept that the
scatter in the stellar to halo mass relation contains significant
information for constraining the physics of galaxy formation. The
simplified models presented here are sufficient to test simple models
of galaxy formation and evolution, and as we isolate the region of
parameter space that is consistent with observations, more
sophistication can be added to these models to properly explore this
parameter space, and more data can be added by measuring clustering of
massive galaxies to test for assembly bias in various physical
quantities. Outside of empirical models of galaxy formation,
semi-analytic and hydrodynamic explorations of galaxy formation
physics should be utilizing $\slogm$ in the assessment of their
models. The processes that regulate star formation will also determine
the scatter in the total amount of star formation. Understanding why
$\slogm$ is so small may be key to our understanding of how the
present day galaxy population came to be.


\bibliography{../risa}

\label{lastpage}

\end{document}